\documentclass[a4paper,11pt,oneside]{article}

\usepackage[english]{babel}
\usepackage{t1enc}% for automatic hyphenation of accented chars
\usepackage[utf8]{inputenc}% for typing chars
\usepackage[T1]{fontenc}
\usepackage{amsmath, amssymb, amsfonts}
\usepackage{indentfirst}
\usepackage{graphicx}
\usepackage{subfig}
\usepackage{color}
\usepackage{amsfonts}
\usepackage{array}
\newcolumntype{L}[1]{>{\raggedright\let\newline\\\arraybackslash\hspace{0pt}}m{#1}}
\newcolumntype{C}[1]{>{\centering\let\newline\\\arraybackslash\hspace{0pt}}m{#1}}
\newcolumntype{R}[1]{>{\raggedleft\let\newline\\\arraybackslash\hspace{0pt}}m{#1}}

\usepackage{anysize}
\marginsize{3cm}{3cm}{2cm}{2cm}
\newcommand{\nc}{\newcommand} % a newcommand r�vid�t�se
\nc{\td}{\mathrm{d}} % �ll� d a deriv�lthoz
\nc{\kd}{\mathbf{k}}
\nc{\Kd}{\mathbf{K}}
\nc{\qd}{\mathbf{q}}
\nc{\Rd}{\mathbf{R}}
\nc{\rd}{\mathbf{r}}
\nc{\ud}{\mathbf{u}}
\nc{\vd}{\mathbf{v}}
\nc{\pd}{\mathbf{p}}
\nc{\Pd}{\mathbf{P}}
\nc{\Gd}{\mathbf{G}}
\nc{\Ad}{\mathbf{A}}
\nc{\Dd}{\mathbf{D}}
\nc{\Sd}{\mathbf{S}}
\nc{\md}{\mathbf{m}}
\nc{\Fd}{\mbox{\boldmath $F$}}
\nc{\Md}{\mbox{$\mathcal{M}$}}
\nc{\bd}{\mbox{\boldmath $\beta$}}
\nc{\Od}{\mbox{$\mathbf{\mathcal{O}}$}}
\nc{\od}{\mbox{\boldmath $\omega$}}
\nc{\Ddd}{\mbox{\boldmath $\underline{\underline{D}}$}}
\nc{\Mdd}{\mbox{\boldmath $\underline{\underline{M}}$}}
\nc{\Cd}{\mbox{\boldmath $\underline{\underline{C}}$}}
\nc{\Div}{\mathrm{div}}
\nc{\Rot}{\mathrm{rot}}
\nc{\Grad}{\mathrm{grad}}
\nc{\Det}{\mathrm{det}}

\title{
	\includegraphics[width=0.35\textwidth]{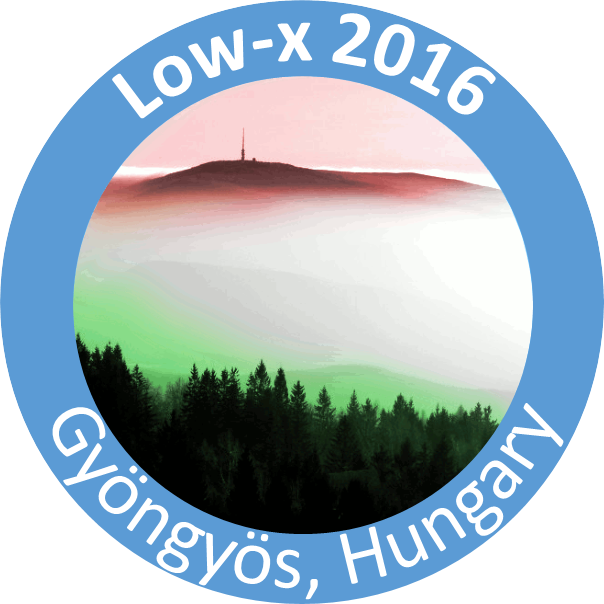}\\[1cm]
	\textbf{New exact solutions of hydrodynamics for rehadronizing fireballs \\ with lattice QCD equation of state}}
\author{{T. Csörgő$^{1,2}$ and  \underline{G. Kasza}$^3$,}\\[1ex]
	$^1$Wigner RCP, H - 1525 Budapest 114, P.O.Box 49, Hungary,\\
	$^2$EKU KRC, H-3200 Gyöngyös, Mátrai út 36, Hungary,\\
	$^3$Eötvös Loránd University, H-1117 Budapest, Pázmány P. s. 1/A, Hungary\\
}
\begin{document}

\maketitle

\begin{abstract} 

We describe fireballs that rehadronize from a perfectly fluid quark matter to
a chemically frozen, multi-component hadron gas.  In the  hydrodynamics of
these fireballs, we utilize the lattice QCD equation of state, however, we
also apply non-relativistic kinematics for simplicity and clarity.  Two new
classes of exact, analytic solutions of fireball hydrodynamics  are presented:
the first class describes triaxially expanding, non-rotating ellipsoidal
fireballs, while the second class of exact solutions corresponds to
spheroidally symmetric, rotating fireballs.  In both classes of solutions, we
find evidence for a secondary explosion, that happens just after hadrochemical
freeze-out.  A realistic, linear mass scaling of the slope parameters of the
single particle spectra of various hadronic species  is obtained analytically,
as well as an also realistic, linear mass scaling of the inverse of the
squared  HBT radius parameters of the Bose-Einstein correlation functions.

\end{abstract}

\section{Introduction}

The equations of hydrodynamics contain no internal scale, and the
applications of hydrodynamics range from the largest, cosmological distances
to the smallest experimentally accessible distances. Hydrodynamical type of equations
characterize the time evolution of our Universe that started from 
a Big Bang. Hydrodynamics is also applied to the study of the time evolution 
on the smallest, femtometer distances, where the Little Bangs of high energy heavy ion collisions 
also create hydrodynamically evolving fireballs.  
Our Universe about 14 billion years after the Big Bang expands with an approximately spherically
symmetric  Hubble flow. The hadronic final states of heavy ion collisions
about a few times 10$^{-23}$ sec after the Little Bangs expand with  
directional Hubble flows and possibly also with significant angular momentum, due to the typically
non-central nature of high energy heavy ion  collisions.

As early as in 1978, Zim\'anyi, Bondorf and Garpman found an exact solution  of
hydrodynamics that described a  non-relativistic, finite fireball with a
Hubble flow, expanding to vacuum~\cite{Bondorf:1978kz}.  
%This solution was first applied to model the radial expansion of the fluid of
%nucleons in intermediate energy heavy ion collisions, but by now this also is
%considered as a first prototype of exact solutions of finite, strongly
%interacting fireballs that model the time evolution of the strongly
%interacting quark gluon plasma in high energy heavy ion collisions.  Several
%such solutions were found recently, some in the non-relativistic, others in
%the relativistic  kinematic domain.  For simplicity, in the present paper we
%focus on the non-relativistic class of solutions.
Keeping the spherical symmetry and the Hubble flow profile, the
Zim\'anyi-Bondorf-Garpman solution was generalized in 1998, after 20 years, to
a spatially Gaussian density and a spatially homogeneous temperature profile,
while maintaining the same equations for the time evolution of the scales as in
the Zimányi-Bondorf-Garpman solution~\cite{Csizmadia:1998ef}.  Soon it was
realized that these solutions can be generalized to arbitrary, but matching
temperature and density profile functions, while still maintaining spherical
~\cite{Csorgo:1998yk} symmetry. Within a few years,  the first, spherically symmetric
solutions were successfully generalized to include  ellipsoidal symmetries
~\cite{Akkelin:2000ex,Csorgo:2001ru}.  About at the same time, the Gaussian
solutions were utilized to evaluate the final state hadronic observables and
their relation to the initial conditions, as it turned out that these solutions
provided exact results for the single particle spectra, elliptic  and higher
order flows, as well as for the Bose-Einstein correlation
functions~\cite{Csorgo:2001xm}.  In this class of solutions, a non-vanishing
initial angular momentum and the corresponding rotation of the expanding
fireball can also be taken into account analytically. The first exact solution
of rotating fireball hydrodynamics was found in the relativistic kinematic
region ~\cite{Nagy:2007xn}. This spheroidally symmetric, relativistic rotating solution was subsequently
generalized to the non-relativistic kinematic domain 
~\cite{Csorgo:2013ksa,Csorgo:2015scx,Nagy:2016uiz}, including not only spheroidally 
but also triaxially expanding and  rotating solutions of fireball hydrodynamics.
In these solutions, the hadronic final state was typically containing only a given
type of particle with mass $m$, and the observables like the slope parameters
of the single particle spectra were investigated as a function of this mass,
considered to be a parameter of the solution.

 This conference presentation details the first steps towards generalizing some
of the recently found expanding as well as rotating, spheroidally and
ellipsoidally symmetric solutions of fireball hydrodynamics
~\cite{Csorgo:2013ksa,Csorgo:2015scx,Nagy:2016uiz} to a more realistic
hadro-chemical and kinetic freeze-out stage. These final states contain a mixture of
hadrons, with different hadronic masses denoted as $m_i$. In this work, we explore two classes of
exact solutions.  The first class describes triaxially expanding, non-rotating
ellipsoidal fireballs, the second class of exact solutions corresponds to
spheroidally symmetric, rotating fireballs.  In both classes of exact
solutions, lattice QCD calculations provide the data for the equations of
state. This allows us to take into account the temperature dependence of the
speed of sound, following refs. ~\cite{Csorgo:2001xm,Csanad:2012hr}.  After
rehadronization, a subsequent hadrochemical freeze-out is shown to have a
significant effect on the expansion dynamics, corresponding to a secondary
explosion, which is seen in  in both classes of exact solutions.  The
properties and the criteria of such a secondary explosion are clarified here in
an exact and analytic manner.

\section{Perfect fluid hydrodynamics for two different stages}

Experimental results of the NA44 ~\cite{Bearden:1996dd} 
as well as the
PHENIX collaborations ~\cite{Adler:2003cb} indicate, for example, that the
so called inverse slope parameter of the single particle spectra is a linear
function of the mass $m$ of the observed hadrons: 
\begin{equation} 
T=T_f+m\langle u_t \rangle ^2, 
\end{equation} 
where $\langle u_t \rangle$	stands for the average radial flow and the
freeze-out temperature is denoted by $T_f$.  This relationship has been derived
even for non-central heavy ion collisions in ref.~\cite{Csorgo:2001xm},
taking into account a chemically frozen, {\it single component} hadronic matter (HM).
However, the experimental data were taken in heavy ion collisions where actually 
several different kind of hadrons are produced simultaneously.
If we introduce an index "$i$" to
distinguish the different  particle types in  a multi-component hadron gas,
then the experimental data indicate that the slope parameters depend on 
the particle type only through the mass $m_i$ of particle type $i$,
but the radial flow $\langle u_t \rangle$ and the kinetic freeze-out temperature
$T_f$ are both independent of the type of the particles: 
\begin{equation} 
T_i=T_{f}+m_i\langle u_t \rangle ^2. 
\end{equation}
In this work, we analytically derive these relations, for a fireball of 
a strongly interacting Quark Gluon Plasma that hadronizes to a {\it multi-component,}
chemically frozen hadronic matter or HM.

The basic equations of perfect fluid hydrodynamics are given by the continuity
and the Euler equation together with the energy equation, corresponding to
local conservation of entropy, momentum and energy. In the strongly coupled
Quark-Gluon Plasma, also called as perfect fluid of Quark Matter or QM, and at
vanishing baryochemical potential, the number of quarks, anti-quarks and gluons
is not conserved individually, only the local conservation of entropy drives
the expansion. However, at a certain temperature various hadrons are produced
due to rehadronization from a QM and we assume in this manuscript that the
inelastic reactions that may transform one hadron to another are negligible, so
we study here the scenario that corresponds to a chemically frozen,
multi-component Hadronic Matter (HM). In this chemically frozen,
multi-component HM phase the number of each type of hadrons is locally
conserved.

The equations of motion for these two different forms of matter 
are summarized in Table ~\ref{table_1}.
These equations generalize the equations of motion for a 
chemically frozen, {\it single component} hadronic matter
equations of (13-16) of ref.~\cite{Csorgo:2013ksa} 
to the case of the chemically frozen, {\it multi-component} scenario of HM. 
%and correspond to the 
%conservation entropy in the sQGP phase. 
%After hadronization, the chemical composition of the
%multi-component hadronic matter (HM) is conserved.
The local momentum and energy conservation, as well as the entropy conservation
is valid in both phases, but in the HM phase, local continuity equations are also
obeyed for all hadronic species.
We utilize the $\epsilon = \kappa p$ equation of state (EoS), 
where $\kappa \equiv \kappa(T) $ is a temperature dependent function, 
that is directly taken from lattice QCD calculations of ref.~\cite{Borsanyi:2010cj}.
We note that in Table ~\ref{table_1} the energy equations are rewritten for the temperature field.
We also note that due to the additional  local conservation laws in the HM phase
the coefficient of the co-moving time derivative of the temperature field changes in the temperature equation
in Table~\ref{table_1}. 
It turnes out that this leads to a dynamical effect, a modification for the time evolution of the temperature.
This modification corresponds to  a secondary explosion that starts at the chemical freeze-out temperature $T_{chem}$.

\begin{table}[ht]
	\centering
	\renewcommand{\arraystretch}{1.5}
	\begin{tabular}{|C {6.7cm}|C {6.7cm}|} 
\hline
		\textbf{QM} $(T_i \ge T \ge T_{chem})$ & \textbf{HM}  $ (T_{chem} > T \ge T_f)$ \\
		\hline 
${\partial_t \sigma}+\nabla\left(\sigma\textbf{v}\right)=0$ & 
${\partial_t n_i}+\nabla\left(n_i \textbf{v}\right)=0,\:\:\:\forall i$ \\
$T\sigma\left(\partial_t+\textbf{v}\nabla\right)\textbf{v}=-\nabla p$ & 
$\sum\limits_i m_i n_i\left(\partial_t+ \textbf{v}\nabla \right)\textbf{v}=
%-T\sum\limits_i \nabla  n_i$ \\ 
-\nabla  p$ \\ 
$\frac{1+\kappa}{T}\left[ \frac{d}{dT} \frac {\kappa T}{1+\kappa} 
\right]\left(\partial_t+\textbf{v}\nabla\right)T+\nabla \textbf{v}=0$ &
 $\frac{1}{T}\left[\frac{d (\kappa T)}{dT}\right]\left(\partial_t + \textbf{v}\nabla\right)T+\nabla\textbf{v}=0$ \\ 		
$p = \sigma T / ( 1 +   \kappa) $ &
$p = \sum\limits_i p_i = T \sum\limits_i n_i$ \\ \hline
	\end{tabular}
	\caption{
Hydrodynamical equations for strongly interacting Quark Gluon Plasma or Quark
Matter (QM) and chemically frozen, multi-component Hadronic Matter (HM) that
drive the fireball expansion from the initial temperature $T_i$ to the chemical
freeze-out temperature ($ T_i \ge T \ge T_{chem}$).  This chemical freeze-out
temperature $T_{chem}$ characterizes both hadronization and simultaneous
hadrochemical freeze-out in the present manuscript.  Below this chemical
freeze-out temperature but above the kinetic freeze-out temperature ($ T_{chem}
> T \ge T_f$), a multi-component hadronic matter is characterized by local
conservation laws for each hadronic species.
}
	\label{table_1}	
\end{table}
In Table ~\ref{table_1}, $ \sigma \equiv \sigma(\textbf{r},t)$ stands for the entropy density,
$n_i \equiv n_i(\textbf{r},t)$ is the density of hadron type $i$ that is locally conserved in the HM phase,
the velocity field is denoted by $\textbf{v} \equiv \textbf{v}(\textbf{r},t)$, 
while $p \equiv p(\textbf{r},t)$ stands for the pressure,
and $T \equiv T(\textbf{r},t)$ for the temperature field, and the mass of hadron type $i$ is denoted as $m_i$. 

As discussed in ref.~\cite{Csorgo:2013ksa}, these equations were derived in the non-relativistic limit of 
the equations of relativistic hydrodynamics, assuming that the entalphy density
(that characterizes the inertia of the motion for pressure gradients) 
is dominated by the entropy density above the chemical freeze-out temperature, while
it is dominated by the mass terms of the hadrons at lower temperatures:
	\begin{eqnarray}
	\varepsilon+p & = & \sum\limits_i \mu_i n_i + T\sigma , \\
	\varepsilon+p & \approx & 
	 T\sigma , \qquad \quad\quad (T_i \ge T \ge T_{chem}), \\
	\varepsilon+p & \approx & 
	\sum\limits_i m_i n_i \qquad ( T_{chem} > T \ge T_f) .
	\end{eqnarray}
The dynamical equations, summarized in Table~\ref{table_1}, can be solved if the usual initial and freeze-out conditions
as well as the chemical freeze-out conditions are given.   In the present work, we characterize these
conditions by the initial temperature $T_i$, the chemical freeze-out temperature $T_{chem}$ and by the kinetic freeze-out temperature $T_f$.

In this manuscript, we also assume that the initial temperature distribution is locally homogeneous,
and its value is given by a coordinate independent $T_i$ value at the initial time $t_i = 0$,
and we also assume that the HM freezes out at a locally homogeneous freeze-out temperature $T_f $. 
In addition to these usual initial and final boundary conditions, 
in these solutions we also have to specify a matching boundary conditions
that specifies the transition from QM to HM, which we characterize by  the locally homogeneous chemical freeze-out temperature $T_{chem}$.

We suppose that rehadronization happens almost simultaneously with the hadrochemical freeze-out at the time $t=t_c$,
and at this temperature the local velocity fields transfer smoothly:
	\begin{eqnarray}
	T_{{B}}(t_c) & = & T_{{A}}(t_c)  \, =\,  T_{{chem}}, \\
        \textbf{v}_{{{B}}}(t_c,\textbf{r})  & = & \textbf{v}_{{A}}(t_c,\textbf{r}). 
	\end{eqnarray}

The medium before the rehadronization is in the QM phase, its parameters are indicated by $B$ that stands for Before.
After the rehadronization, we use the $A$ index, it indexes the medium that is converted to the HM phase. 
We follow Landau's proposal, who suggested that at the time of rehadronization
a conversion takes place between entropy density and particle density~\cite{Belenkij:1956cd}. 
Therefore we assume that 
	\begin{equation}
	\frac{\sigma(\textbf{r},t_c)}{\sigma(\textbf{r} = 0, t_c)}=\frac{n_i(\textbf{r},t)}{n_{i}(\textbf{r}=0,t_c)}.
	\end{equation}

We look for parametric solutions of the hydrodynamical equations, summarized in Table~\ref{table_1},
and we assume that the principal axes of a triaxially expanding, ellipsoidal  fireball
are  be given by $X \equiv X(t)$, $Y \equiv Y(t)$ and $Z \equiv Z(t)$ that functions depend only on the time $t$.

In this manuscript, we discuss two classes of parametric, exact solutions of fireball hydrodynamics.
The first class is a triaxial, non-rotating class of solutions, while the second class corresponds to  a spheroidally symmetric, rotating class
of exact solutions of fireball hydrodynamics. In the triaxial case, all the principal axis $(X,Y,Z)$ can be different,
but the initial angular velocity $\omega_0$ has to vanish.  For the rotating solutions of fireball hydrodynamics 
with non-vanishing initial angular velocity,  
we assume spheroidal symmmetry and introduce the notation $X(t)=Y(t)=R(t)$.

All of the scale functions $(X,Y,Z)$ as well as $R$ are continuous at $t_c$,
and it turns out that we can follow the lines of derivations described in
refs.~\cite{Csorgo:2001ru,Csorgo:2013ksa,Csorgo:2015scx} even for an QM that
rehadronizes to a HM, without introducing a particle species dependence of
the scale parameters $(X,Y,Z)$ after the rehadronization. The details of these
calculations are not given here, but the main results are summarized in
Table ~\ref{table_2a} for a tri-axially expanding, non-rotating ellipsoidal fireball,
and  Table ~\ref{table_2b} for a spheroidal, rotating and expanding fireball.
These results indicate that the rather complicated partial
differential equations that govern the dynamics of the fireball expansion can
be solved exactly, when the hydrodynamical fields are given in terms of the scale parameters
of the solutions. Thus these solutions are parametric solutions, the scale
parameters $(X,Y,Z)$  satisfy a system of coupled and non-linear but
ordinary differential equations, listed also in Tables ~\ref{table_2a} and \ref{table_2b}. 
These differential equations can be readily solved with currently available numerical packages like { \sc
Mathematica} or {\sc Matlab }.

\begin{table}
\centering	
	\renewcommand{\arraystretch}{1.5}
\begin{tabular}{|c|c|}			
\hline
 \textbf{QM} $(T_i \ge T \ge T_{chem})$ & \textbf{HM}  $ (T_{chem} > T \ge T_f)$ \\ \hline 
$ {\textbf v} = (\frac{\dot X}{X} r_x, \frac{\dot Y}{Y} r_y, \frac{\dot Z}{Z} r_z)$ &
$ {\textbf v} = (\frac{\dot X}{X} r_x, \frac{\dot Y}{Y} r_y, \frac{\dot Z}{Z} r_z)$ \\ 
$ \sigma = \sigma_0 \frac{V_0}{V} \exp\left(- \frac{r_x^2}{2 X^2} - \frac{r_y^2}{2 Y^2} -\frac{r_z^2}{2 Z^2}\right)$	& 
		$ n_i = n_{i,c} \frac{V_c}{V} \exp\left(- \frac{r_x^2}{2 X^2} - \frac{r_y^2}{2 Y^2} -\frac{r_z^2}{2 Z^2}\right)$	 	 \\ \hline
$\left(1+\kappa\right)\left[ \frac{d}{dT} \frac {\kappa T}{1+\kappa} \right]\frac{\dot{T}}{T}+\frac{\dot{V}}{V}=0$ & 
			$\frac{d\left(\kappa T\right)}{dT}\frac{\dot{T}}{T}+\frac{\dot{V}}{V}=0$ \\
$X\ddot{X}=Y\ddot{Y}=Z\ddot{Z}=\frac{1}{1+\kappa(T)}$ & $X\ddot{X}=Y\ddot{Y}=Z\ddot{Z}=\frac{T}{\langle m \rangle}$ \\ \hline
\end{tabular}
\caption{Parametric solution of fireball hydrodynamics for a tri-axially
expanding, non-rotating  ellipsoidal fireball, 
where the volume $V$ and the
average mass  $\langle m \rangle$ are defined by eqs. (\ref{e:volume}) and
(\ref{e:averagem}).  
The first two rows give the parametric form of the density and the velocity
fields. Note that  in these solutions, the  corresponding temperature field is homogeneous,
$T(t,\textbf r ) \equiv T(t)$. The time evolution of the temperature is
determined by an ordinary differential equation, that depends on the Equation of
State through the function $\kappa \equiv \kappa(T)$ which for a
spatially homogeneous temperature field is a function of time only, $\kappa \equiv
\kappa(T(t))$.  The acceleration of the scales $X,Y,Z$ is driven also by the equation of state,
but on the QM side the value of the constant of proportionality, $\frac{1}{1 + \kappa(T)} $ is in general different
from the value of constant of proportionality in the HM phase, $\frac{T}{\langle m \rangle}$.
}
 \label{table_2a}
\end{table}

\begin{table}
\centering	
\renewcommand{\arraystretch}{1.5}
\begin{tabular}{|c|c|}			
\hline 
 \textbf{QM} $(T_i \ge T \ge T_{chem})$ & \textbf{HM}  $ (T_{chem} > T \ge T_f)$ \\ \hline 
$ {\textbf v} = (\frac{\dot R}{R} r_x - \omega r_y, \frac{\dot R}{R} r_y + \omega r_x, \frac{\dot Z}{Z} r_z)$ &
$ {\textbf v} = (\frac{\dot R}{R} r_x - \omega r_y, \frac{\dot R}{R} r_y + \omega r_x, \frac{\dot Z}{Z} r_z)$ \\
$ \sigma = \sigma_0 \frac{V_0}{V} \exp\left(- \frac{r_x^2}{2 R^2} - \frac{r_y^2}{2 R^2} -\frac{r_z^2}{2 Z^2}\right)$	& 
$ n_i = n_{i,c} \frac{V_c}{V} \exp\left(- \frac{r_x^2}{2 R^2} - \frac{r_y^2}{2 R^2} -\frac{r_z^2}{2 Z^2}\right)$ \\ \hline 
 $\left(1+\kappa\right)\left[ \frac{d}{dT} \frac {\kappa T}{1+\kappa} \right]\frac{\dot{T}}{T}+\frac{\dot{V}}{V}=0$ & $\frac{d\left(\kappa T\right)}{dT}\frac{\dot{T}}{T}+\frac{\dot{V}}{V}=0$ \\ 
 $R\ddot{R}-R^2\omega^2=Z\ddot{Z}=\frac{1}{1+\kappa(T)}$  &$R\ddot{R}-R^2\omega^2=Z\ddot{Z}=\frac{T}{\langle m \rangle}$ \\ \hline
	\end{tabular}
\caption{Parametric solution of fireball hydrodynamics for a spheroidally 
expanding, and  rotating  fireball. Notation is the similar to that of Table ~\ref{table_2a},
but the volume $V$ is defined by  eq. (\ref{e:volume-spheroid}) and the time evolution of the angular velocity $\omega$
is given by eq. ~(\ref{e:omega}).  
}
	\label{table_2b}
\end{table}
For a triaxially expanding ellipsoid, the volume of the fireball is given by that of a 3d Gaussian with widths $X$, $Y$ and $Z$:
\begin{equation}
V(t) =(2 \pi)^{3/2} X Y Z ,
\label{e:volume}
\end{equation}
while for a spheroidally expanding ellipsoid, $X = Y = R$ and the volume is given by 
\begin{equation}
V(t) =(2 \pi)^{3/2} R^2 Z .
\label{e:volume-spheroid}
\end{equation}
In the considered class of exact, rotating spheroidal solution the angular velocity is driven by the
radial expansion as follows:
\begin{equation}
\omega(t)=\omega_0\frac{R_0^2}{R(t)^2}.
\label{e:omega}
\end{equation}

In this expression $\omega_0$ and $R_0$ are the initial values of the
corresponding functions at the initial time $t_0$.  As the equations of motion
for the scales are indepenent from the type of particle $i$ in the HM
phase, it is easy to see 
that the fireball expands collectively to the vacuum, for all particle types $i$.

Instead of the mass $m$ of a single type of particle in the dynamical equations of  
a {\it single component}, chemically frozen HM phase, the average mass $\langle m\rangle$
appears in the dynamics of a  {\it multi-component}, chemically frozen HM phase.
The typical value for $\langle m\rangle$ 200 GeV Au+Au collisions
at RHIC is given approximately~\cite{Kaneta:2004zr}
 as
\begin{equation}
\langle m\rangle=\frac{\sum\limits_i m_i n_{i,c}}{\sum\limits_i n_{i,c}}\approx 280\:MeV.
\label{e:averagem}
\end{equation}				
The same analysis~\cite{Kaneta:2004zr} indicated chemical freeze-out temperatures in the range of $T_{chem} ~ 150-170 $ MeV.
At the chemical freeze-out ($T\approx T_{chem}$), the acceleration changes
due to the change of the coefficients that determine $\ddot X X$ and similar quantities.
To quantify this, we evaluate the right hand side of the acceleration equations at $T_{chem}$,
both in the QM and in the HM phases, using the lattice QCD equation of state, 
and we find the following relation: 
\begin{equation}
	\frac{1}{1+\kappa(T_{chem})} \simeq 0.11 - 0.15 < \frac{T_{chem}}{\langle m \rangle} \simeq 0.55 - 0.63 .
	\label{e:explosion}
\end{equation}
This inequality is thus valid in  a broad range of $T_{chem}$, independently from the actual value of the
chemical freeze-out temperature, if this is varied in the reasonable range of
150 $< T_{chem} < $ 175 MeV~\cite{Kaneta:2004zr}.

As a consequence, the acceleration of the scales $(X,Y, Z)$ starts to {\it
increase} as the temperature cools just below $T_{chem}$, for any reasonable value of $T_{chem}$,
not due to the change of the pressure but due to the change of the dynamical equations,
that include new conservation laws.  
This increased acceleration leads to a secondary explosion of the medium, which starts just after
the conversion from quark matter to the chemically frozen hadronic matter.  

A novel feature of the secondary explosion is that actually this happens at
temperatures  where the $\kappa = \varepsilon/p$ ratio is close to its maximum
in lattice QCD calculations, hence the corresponding speed of sound is nearly
minimal. This temperature is usually called the ``softest point" of the equation
of state, and it is usually associated with a slowing down of the transverse flows,
see for example the exact solutions of T. S. Bir\'o for a first order phase transition
of a massless gas of quarks and  gluons to a massless pion gas ~\cite{Biro:1999eh,Biro:2000nj}. 
 In particular if the pressure could become a constant during a first order phase transition,
its gradiends would approach vanishing values, hence the acceleration terms would vanish.  
However, when we take into account  a lattice QCD equation of state, that lacks a first order phase 
transition at small baryochemical potentials, the pressure gradients 
do not vanish. Furthermore, at $T_{chem}$, additional local hadronic  conservation laws start to play
a role and modify the dynamics. As a consequence of inequality
in eq. ~(\ref{e:explosion}), instead of slowing
down, the expansion starts actually  to accelerate faster at $T_{chem}$, as compared to the case when
hadronization and hadrochemical freeze-out does not happen!

Another novel and rather surprising feature of this secondary explosion is related to the relative
position of the chemical freeze-out to the softest point of the lQCD Equation
of State. If the chemical freeze-out temperature $T_{chem}$ is less than
$T_{max} \approx 151 $ MeV, the temperature where $d\kappa/dT(T = T_{max}) =
0$, this second explosion generated by the hadrochemical freeze-out leads to
{\it faster expansion} as well as {\it slower cooling}, as compared to an
expansion where hadrochemical freeze-out does not happen.  This is a rather
unusual scenario, as normally faster expansion leads to faster cooling. Such a
more usual behaviour is described by the same equations if $T_{chem} > 
T_{max} = 151 $ MeV. As this is the expected range for the chemical freeze-out
temperatures~\cite{Kaneta:2004zr},
we expect that when the secondary, hadrochemical explosion happens, and the
fireball starts to expand faster, the cooling of the temperature as a function
of time actually becomes also faster. 

\section{Observables}

The observables for  a single-component hadronic matter (HM) were already
evaluated in refs.~\cite{Akkelin:2000ex} and \cite{Csorgo:2015scx}. In this
manuscript we  present the generalization of these earlier results for the
multi-component hadronic matter scenario.  The results are summmarized in
Tables ~\ref{table_3a} and ~\ref{table_3b}, corresponding to the solutions in
Table~\ref{table_2a} and ~\ref{table_2b}, respectively.  These results
summarize only some of the key, the selected hadronic observables, such as the
inverse slope parameters and the HBT-radii.  The relation of these key
observables to the single particle spectra, elliptic or higher order flows or
to the Bose-Einstein correlation functions is the same, as in
refs.~\cite{Csorgo:2001xm,Csorgo:2015scx}, respectively.  In these
calculations, the freeze-out temperature is denoted by $T_f$ and subscript $f$
indicates quantities that are evaluated at the time of the kinetic feeze-out.	
	
\begin{table}[ht]		
\centering
\renewcommand{\arraystretch}{1.3}
\begin{tabular}{|C{5.9cm}|C{5.9cm}|}			
\hline
	\textbf{HM} (one kind of hadron only, with mass $m$)& \textbf{HM} (mixture of various hadrons, with masses $m_i$)\\ \hline 
	$T_x=T_f+m \dot{X_f}^2$ & $T_{x,i}=T_f+m_i \dot{X_f}^2$ \\
	$T_y=T_f+m \dot{Y_f}^2$ & $T_{y,i}=T_f+m_i \dot{Y_f}^2$ \\
	$T_z=T_f+m \dot{Z_f}^2$ & $T_{z,i}=T_f+m_i \dot{Z_f}^2$\\ \hline
	$R_{x}^{-2}=X_f^{-2}\left[1+\frac{m}{T_f}\dot{X}_f^2\right]$ & $R_{x,i}^{-2}=X_f^{-2}\left[1+\frac{m_i}{T_f}\dot{X}_f^2\right]$ \\
	$R_{y}^{-2}=Y_f^{-2}\left[1+\frac{m}{T_f}\dot{Y}_f^2\right]$ & $R_{y,i}^{-2}=Y_f^{-2}\left[1+\frac{m_i}{T_f}\dot{Y}_f^2\right]$\\
	$R_{z}^{-2}=Z_f^{-2}\left[1+\frac{m}{T_f}\dot{Z}_f^2\right]$ & $R_{z,i}^{-2}=Z_f^{-2}\left[1+\frac{m_i}{T_f}\dot{Z}_f^2\right]$\\ \hline
\end{tabular}
\caption{
Inverse slope parameters for a single component and a multi-component hadronic
matter as well as HBT-radii  for a
triaxially expanding, non-rotating, ellipsoidal fireball, corresponding
to the hydrodynamical solution in Table~\ref{table_2a}. 
The relation to the single particle spectra and Bose-Einstein correlation functions
is the same, as in ref.~\cite{Csorgo:2001xm}, but instead of the mass $m$ of a single kind of hadron
for each hadronic species $i$ their mass $m_i$ appears in the observables.
}
\label{table_3a}	
\end{table}

\begin{table}[ht]		
\centering
\renewcommand{\arraystretch}{1.3}
\begin{tabular}{|C{5.9cm}|C{5.9cm}|}			
\hline
	\textbf{HM} (single component, with mass $m$)& \textbf{HM} (multi-component, with masses $m_i$)\\ \hline 
	$T_x=T_f+m\left(\dot{R_f}^2+\omega_f^2 R_f^2\right)$ & $T_{x,i}=T_f+m_i\left(\dot{R_f}^2+\omega_f^2 R_f^2\right)$ \\
	$T_y=T_f+m\left(\dot{R_f}^2+\omega_f^2 R_f^2\right)$ & $T_{y,i}=T_f+m_i\left(\dot{R_f}^2+\omega_f^2 R_f^2\right)$ \\
	$T_z=T_f+m \dot{Z_f}^2$ & $T_{z,i}=T_f+m_i \dot{Z_f}^2$ \\ \hline
	$R_{x}^{-2}=R_f^{-2}\left[1+\frac{m}{T_f}\left(\dot{R}_f^2+R_f^2\omega_f^2\right)\right]$ & $R_{x,i}^{-2}=R_f^{-2}\left[1+\frac{m_i}{T_f}\left(\dot{R}_f^2+R_f^2\omega_f^2\right)\right]$ \\
	$R_{y}^{-2}=R_f^{-2}\left[1+\frac{m}{T_f}\left(\dot{R}_f^2+R_f^2\omega_f^2\right)\right]$ & $R_{y,i}^{-2}=R_f^{-2}\left[1+\frac{m_i}{T_f}\left(\dot{R}_f^2+R_f^2\omega_f^2\right)\right]$\\
	$R_{z}^{-2}=Z_f^{-2}\left[1+\frac{m}{T_f}\dot{Z}_f^2\right]$ & $R_{z,i}^{-2}=Z_f^{-2}\left[1+\frac{m_i}{T_f}\dot{Z}_f^2\right]$\\ \hline
\end{tabular}
\caption{
Inverse slope parameters for a single component and a multi-component hadronic
matter as well as HBT-radii for a rotating and expanding spheroidal fireball, corresponding
to the hydrodynamical solution in Table~\ref{table_2b}. The 
relation to the single particle spectra and Bose-Einstein correlation functions
is the same, as in ref.~\cite{Csorgo:2015scx}, but the results for the single component
hadron mass are generalized for the multi-component scenario.
The new results can be obtained simply, with the help of an $m \rightarrow m_i$
replacement .
}
\label{table_3b}	
\end{table}
The inverse slopes and the
squared inverse HBT-radii are linear functions of $m_i$. Recent
experimental results of for example the PHENIX collaboration correspond well to
these linear relations~\cite{Kincses:2016mxp}. As these data were taken in high energy heavy ion 
collisions, where the hadronic final state contains a mixture
of various hadrons (referred to as the multi-component Hadronic Matter scenario),
it is a non-trivial result that such simple replacement rules:
$m\rightarrow \langle m\rangle$ in the dynamical equations 
and $m \rightarrow m_i$ in the observables can be utilized to obtain the new exact solutions
of the hydrodynamical equations and the evaluation of the obsverables.

\section{A new parametrization for lattice QCD EoS} 
In the earlier sections of this manuscript we presented the transition of a Quark Matter to Hadronic Matter that contained
a mixture of various hadrons. These solutions, however, were limited by the assumption of a homogeneous initial temperature profile.
In this section we prepare the ground for new solutions where the initial temperature and density profile may be inhomogeneous.

Recently, ref.~\cite{Csorgo:2013ksa}
 explored new, exact, parametric solutions
of non-relativistic, rotating fireballs, using a lattice QCD equation of state,
similarly to our previous studies, but using a single mass $m$ in the hadron
gas phase.  That work explored two kinds of exact solutions: the first class of
solutions had homogeneous temperature profiles, where the local temperature was
a function of time only, $T \equiv T(t)$ .  That class of solutions were
generalized to the multi-component hadronic matter in the previous sections of
this manuscript. The second class of solutions in ref.~\cite{Csorgo:2013ksa}
allowed for inhomogeneous temperature profiles if the density profiles had a
corresponding, matching shape. This second class of solutions was obtained for
a special equation of state, where the $\kappa(T) \equiv \kappa_c$ function was
a temperature independent constant. We are not interested here in this
scenario, as the lattice QCD Equation of State indicates that $\kappa =
\varepsilon/p $ is not a temperature independent constant.  However, in a
footnote of ref.  ~\cite{Csorgo:2013ksa}, a third class of solutions was also
mentioned, noting that solutions exist also for the case of inhomogeneous
temperature profiles also in the case of a temperature dependent $\kappa(T)$
functions, if a special differential equation is statisfied by $\kappa(T)$
functions, however, this class was not investigated in detail. 

Here we follow up that line of research by demonstrating that the lattice QCD
equation of state can be parameterized by $\kappa(T)$ functions that allow for
exact solutions of fireball hydrodynamics with inhomogeneous temperature
profiles.  The criteria to find such hydrodynamical solutions is that the
coefficient of the logarithmic comoving derivative of the temperature fields be
a constant both in the QM and in the HM phase, as detailed below.

From the temperature equation for high temperatures $(T_i \ge T \ge T_{chem})$,
corresponding to the dynamical equations that describe the evolution of QM in
Table~\ref{table_1}, this criteria leads to the following constraint on the
possible shape of the  $\kappa(T) $ function:

\begin{equation}
	\frac{d}{dT}\left[\frac{T\kappa(T)}{1+\kappa(T)} \right]=\frac{\kappa_{Q}}{1+\kappa(T)}, \qquad (T \ge T_{chem}),
\label{e:sQGPconstraint}
\end{equation}

where $\kappa_Q = \lim_{T\rightarrow \infty} \kappa(T)$ stands for the high
temperature limit of the $\kappa(T)$ function.

As the coefficient of the temperature equation in Table~\ref{table_1} is
modified at lower temperatures ($T_{chem}> T > T_f$), 
corresponding to a multi-component, chemically frozen Hadronic Matter, in this
temperature range a modified  constraint is obtained for the $\kappa(T)$
function:

\begin{equation}
	\frac{d}{dT}\left[T\kappa(T)\right]=\frac{\kappa_c T_c-\kappa_f T_f}{T_c-T_f}. \qquad (T_{chem} > T \ge T_f), \label{e:HG-constraint}
\end{equation}

where $T_c=T_{chem}=175\:MeV$ corresponds to the upper limit of
the chemical freeze-out temperatures obtained from  experimental data 
on particle ratios in $\sqrt{s_{NN}} = 200$ GeV Au+Au collisions at
 RHIC~\cite{Kaneta:2004zr}.
In the above equations, we
have assumed that at the kinetic freeze-out the non-relativistic ideal gas
approximation  can be used i.e. $\kappa_f = \kappa(T_f)=3/2$, however higher
values of $\kappa_f$ can also be used if one intends to match lattice QCD calculations
at lower temperatures closely.  In any case, after freeze-out we assume that hadrons propagate
to the detectors with free streaming and post kinetic freeze-out their energy density to pressure
ratio thus decreases or jumps to the value of  3/2.

For the QM phase the analytic solution of the constraint (\ref{e:sQGPconstraint}) is

\begin{equation}
\kappa_{QM}(T)=\frac{\kappa_Q \left(\frac{T}{T_c}\right)^{1+\kappa_Q}+\frac{\kappa_c-\kappa_Q}{\kappa_c+1}}{\left(\frac{T}{T_c}\right)^{1+\kappa_Q}-\frac{\kappa_c-\kappa_Q}{\kappa_c+1}},
\end{equation}		

and in this function $\kappa_c$ stands for $\kappa(T_c)$. For the HM phase, the solution to the constraint of 
eq. (~\ref{e:HG-constraint}) yields the following form for $\kappa(T)$ :
\begin{equation}
\kappa_{HM}(T)=\frac{\kappa_c T_c-\kappa_f T_f}{T_c-T_f}-\frac{\kappa_c-\kappa_f}{T_c-T_f}\frac{T_c T_f}{T}.
\end{equation}

\begin{figure}[ht]
	\centering
	\includegraphics[scale=0.8]{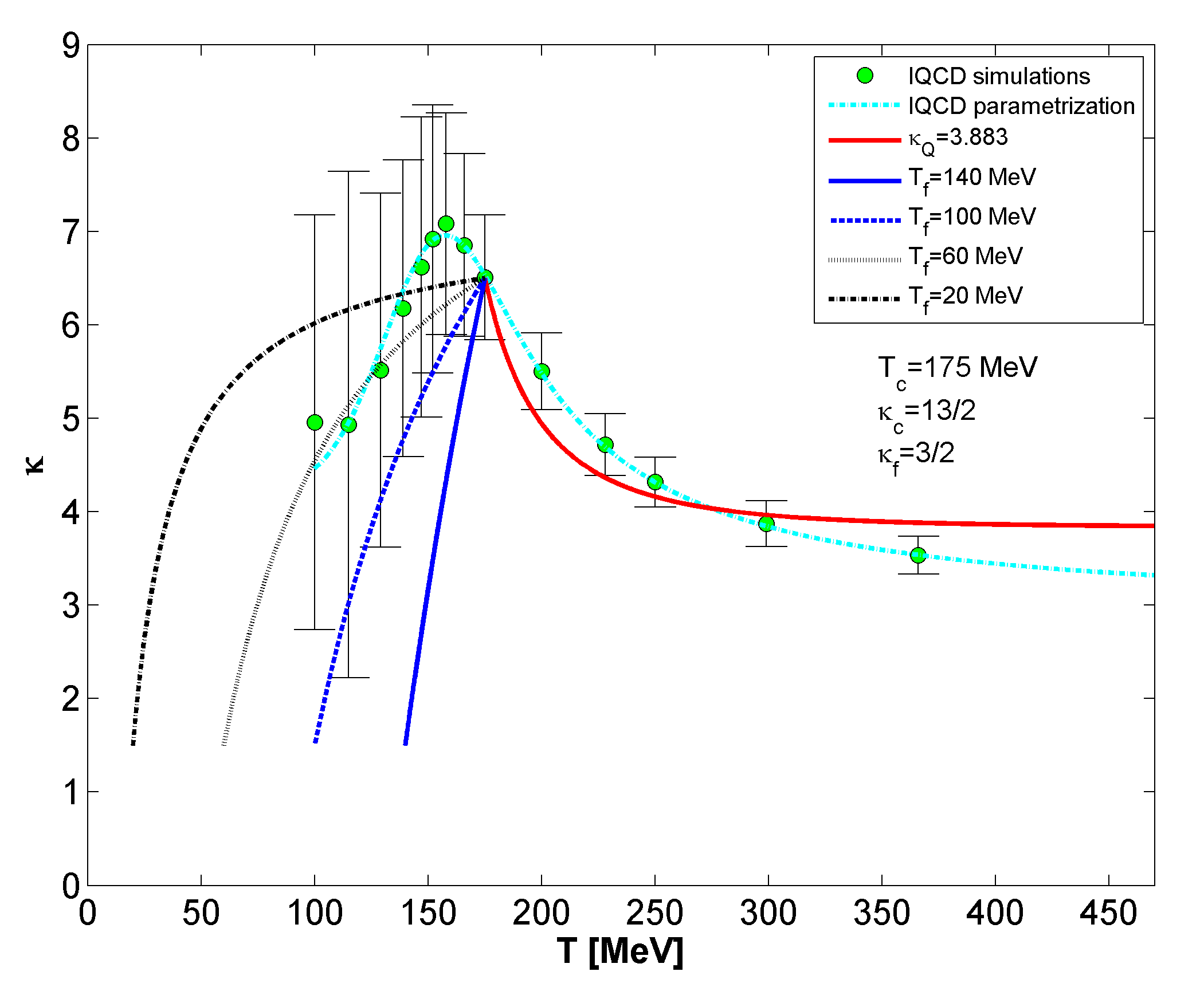}~\\*
	\caption{
Fits of the hydrodynamically motivated parameterizations above and below
$T_{chem}$ to the  lattice QCD data points on $\kappa(T) = \varepsilon/p$.
In these fits, we required
that at freeze-out, a non-relativistic ideal gas limit is approached so that
$\kappa(T_f) = 3/2$ and varied the freeze-out temperature from 20 to 140 MeV.
}
	\label{fig_1}
\end{figure}

These solutions are matched  at the critical temperature $T_c=175$ MeV and we
have assumed that the chemical freeze-out temperature is the same as the
critical temperature, $T_{chem} = T_c$. We made fits to simulated data from
lattice QCD ~\cite{Borsanyi:2010cj} using $\kappa_Q$ as a fitting parameter,
for  $T_c = 175$ MeV fixed and using various values of the kinetic freeze-out
temperature $T_f$.  The quality of these fits is summarized in Table
\ref{table_4} and on Figure \ref{fig_1}. 

In the QM phase, a satisfactory fit is found, as indicated by the red curve and
summarized also in Table \ref{table_4}.  We could also obtain reasonably good
fits in the HM range of temperatures, however, with some constaints on the
possible value of the kinetic freeze-out temperature $T_f$: A reasonable value
of the freeze-out temperature is the pion mass, $T_f \approx 140$ MeV
(continuous, blue line) but in this case $\kappa$ falls down too steeply with
temperature due to our additional requirement of $\kappa(T_f) = 3/2$ and it is
reflected very well by the unsatisfactory confidence level of this fit.
However, fits with freeze-out temperature $T_f \le 100$ MeV and $\kappa(T_f) =
3/2$ are statistically acceptable.

\begin{table}[ht]
	\centering
	\renewcommand{\arraystretch}{1.2}
	\begin{tabular}{|l|c|c|}
		\hline
		Curves & $\chi^2/NDF$ & $CL\:[\%]$\\
		\hline \hline
		lQCD parametrization & 0.12/5 & $>99.9$\\
		$\kappa_Q=3.833$ & 6.48/4 & 16.6\\
		$T_f=140\:MeV$ & 86.56/6 & 1.6$\cdot10^{-14}$\\
		$T_f=100\:MeV$ & 7.71/6 & 26.0\\
	%	$T_f=60\:MeV$ & 1.35/6 & 96.9\\
	%	$T_f=20\:MeV$ & 1.22/6 & 97.6\\
		\hline		
	\end{tabular}
	\caption{Confidence levels of parametrizations of the lattice QCD Equation of State, for various values of the freeze-out temperature $T_f$.
Note that in these parameterizations, $\kappa(T_f) = 3/2$, so at freeze-out a non-interacting, ideal gas equation of state is reached.}	
	\label{table_4}
\end{table}

This section prepares the ground for new exact analytic solutions of hydrodynamics where the initial temperature profile is spatially inhomogeneous.
Although such solutions can be obtained by straight-forward generalizations of the exact solutions of ref. ~\cite{Csorgo:2013ksa}
with spatially inhomogeneous temperature profiles both in the high temperature QM and in the low temperature HM phases, even for a multi-component
hadronic matter scenario, their matching at the chemical freeze-out temperature is an open research question hence these solutions are not detailed here.

\section{Conclusions}

We described two new classes of exact solutions of fireball hydrodynamics, for
a rehadronizing and expanding fireball, using lattice QCD Equation of State.
In the first class of solutions, the expaning ellipsoid is triaxial, but the
fireball is not rotating, ($X \ne Y \ne Z, \omega = 0$).  In the second
class of solutions, although the expansion is spheroidal, 
the fireball is  rotating, ($X = Y = R \ne Z, \omega \ne 0$).  
In both cases, we found that the fireball
expands to the vacuum as a whole, although the quark matter rehadronizes to a
hadronic matter that includes various hadronic components (for example pions,
kaons, protons and all the other measured hadronic species). In both classes of
the presented new solutions, the same length and temperature  scales
characterize the fireball dynamics for all the hadronic types in the final
state, ($ X \ne X_i, Y\ne Y_i, Z \ne Z_i$), so the fireball keeps on expanding
as a whole, instead of developing non-equilibrium features such as separate
lenght-scales for each observable hadrons.

We have obtained a surprising analytic insight to the effects of hadrochemical
freeze-out on the expansion dynamics. If rehadronization is immediately
followed by a  hadrochemical freeze-out, this  leads to a modification of the
dynamical equations, which in turn leads to a  second, violent, hadrochemical
explosion. Instead of slowing down the radial flows at the  softest point where
$p/\varepsilon$ is minimal, the expansion dynamics does not slow down, but it
actually accelerates.  We have found that the expansion dynamics starts to
accelerate at the chemical freeze-out temperature due to the inequality
(\ref{e:explosion}) which is a consequences of the application of lattice QCD
EoS when evaluating the expansion dynamics in Tables ~\ref{table_2a} and
~\ref{table_2b}.  In this hadrochemical explosion, all the length-scales $(X,
Y, Z)$  and $R$ start to accelerate faster, when the temperature drops just
below $T = T_{chem}$, as compared to a scenario without hadrochemical
freeze-out, so in this sense the dynamics becomes ``hardest" at the ``softest point" 
of the lattice QCD Equation of State. 

In the last section, we have also shown that the lattice QCD equation of state
$\kappa(T)$  can be parametrized in  a new way, which is suitable for the
development of exact and analytic, parametric solutions of fireball
hydrodynamics even for an initially inhomogenous temperature profile. The
details of this solution with inhomogeneous temperature profile, as well as the
extension of the presented solutions to the relativistic kinematic region are
important issues that  go beyond the scope of the limitations of this
conference contribution.
	
\section*{Acknowledgments} 
We thank Y. Hatta, D. Klabucar, T. Kunihiro, S. Nagamiya and K. Ozawa for enlightening and useful discussions.  
T. Cs. would like to thank S. Nagamiya and K. Ozawa for their kind hospitality at KEK, Tsukuba, Japan.
This research was supported by the Hungarian OTKA grant NK 101438 as well as by a 2016 KEK Visitor Fund.

\vfill\eject

%\nocite{*}
\bibliographystyle{unsrt}
%\bibliography{cikk}		

\end{document}